\documentclass[preprint]{aastex}               

\received{}
\revised{}
\accepted{}

\cpright{AAS}{2000}

\slugcomment{submitted to ApJL}

\shorttitle{{\sl HST\/}/STIS Imaging of GRB~980425/SN1998bw}
\shortauthors{Fynbo et~al.}

\begin{document}

\title{{\sl HST\/}/STIS Imaging of the Host Galaxy of
       GRB~980425/SN1998bw {}\footnote{\rm Based on observations with
       the NASA/ESA {\sl Hubble Space Telescope\/} ({\sl HST\/}),
       obtained at the Space Telescope Science Institute, which is
       operated by the Association of Universities for Research in
       Astronomy, Inc.\ under NASA contract No.\ NAS5-26555 and on
       observations made with ESO Telescopes at the Paranal
       Observatory under programme ID 63.O-0065.}}

\author{J. U. Fynbo}
\affil{Astronomical Observatory,
       University of Copenhagen,
       Juliane Maries Vej 30,
       DK--2100 Copenhagen {\O},
       Denmark}
\email{jfynbo@ifa.au.dk}

\author{S. Holland}
\affil{Institute of Physics and Astronomy,
       University of Aarhus,
       DK--8000 {\AA}rhus C,
       Denmark}
\email{holland@ifa.au.dk}

\author{M. I. Andersen}
\affil{Division of Astronomy,
       University of Oulu,
       P. O. Box 3000,
       FIN--90014 University of Oulu,
       Finland}
\email{manderse@sun3.oulu.fi}

\author{B. Thomsen}
\affil{Institute of Physics and Astronomy,
       University of Aarhus,
       DK--8000 {\AA}rhus C,
       Denmark}
\email{bt@ifa.au.dk}

\author{J. Hjorth}
\affil{Astronomical Observatory,
       University of Copenhagen,
       Juliane Maries Vej 30,
       DK--2100 Copenhagen {\O},
       Denmark}
\email{jens@astro.ku.dk}

\author{G. Bj{\"o}rnsson}
\affil{Science Institute,
       Dunhagi 3,
       University of Iceland,
       IS--107 Reykjavik,
       Iceland}
\email{gulli@raunvis.hi.is}

\author{A. O. Jaunsen}
\affil{Institute of Theoretical Astrophysics,
       University of Oslo, 
       Blindern,
       N--0315 Oslo,
       Norway}
\email{a.o.jaunsen@astro.uio.no}

\author{P. Natarajan}
\affil{Institute of Astronomy,
       Madingley Road,
       Cambridge CB3 0HA,
       England,
       and
       Department of Astronomy,
       Yale University,
       New Haven, CT 06520--8181,
       U.S.A.}
\email{priya@ast.cam.ac.uk}

\and

\author{N. Tanvir}
\affil{Department of Physical Sciences,
       University of Hertfordshire,
       College Lane, Hatfield,
       Hertfordshire AL10 9AB,
       England}
\email{nrt@ast.cam.ac.uk}


\begin{abstract}
	We present {\sl HST\/}/STIS observations of ESO~184--G82, the
host galaxy of the gamma-ray burst GRB~980425 associated with the
peculiar Type Ic supernova SN1998bw.  ESO~184--G82 is found to be an
actively star forming SBc sub-luminous galaxy. We detect an object
consistent with being a point source within the astrometric
uncertainty of $0\farcs018$ of the position of the supernova.  The
object is located inside a star-forming region. The object is $\gtrsim
1$ magnitude brighter than expected for the supernova based on a
simple radioactive decay model.  This implies either a significant
flattening of the light curve or a contribution from an underlying
star cluster.
\end{abstract}



\section{Introduction}

	There is evidence that at least some gamma-ray bursts (GRBs)
are related to supernovae (SNe). GRB~980425, GRB~970514 and GRB~980919
are consistent with being in temporal as well as spatial coincidence
with SN1998bw, SN1997cy and SN1999E, respectively
\citep{GAL,Ger,Tu,Tho}.  The associations between GRB~970514 and
SN1997cy, and between GRB~980919 and SN1999E, could be by chance.
However, this is very unlikely for GRB~980425 since the temporal and
spatial constraints are much tighter than for the two other possible
GRB/SN associations \citep{GAL}.  Furthermore, there is evidence for a
contribution from SNe to the light-curves of GRB~970228 and GRB~980326
\citep{D1999,Ric,BKD1999,CTG1999,GAL2}.  However, less than 0.2\% of SNe
release detectable gamma-ray emission \citep{Kip,Woo}, and SN1997cy,
SN1998bw and SN1999E were extremely bright and peculiar SNe.  This
association between GRBs and peculiar SNe suggests that GRB~980425
(and possibly also GRB~970514 and GRB~980919) were members of a rare
class of GRBs and may not be indicative of normal GRBs.  However, the
temporal and spatial properties of the gamma radiation from
GRB~980425, except for the low total luminosity, were not unusual
\citep{GAL,PAA2000}.

	In this {\sl Letter\/} we present Space Telescope Imaging
Spectrograph (STIS) clear and long-pass images of the host galaxy of
SN1998bw as part of a large survey aimed at studying the morphology of
GRB host galaxies and the locations of the GRBs within the host
galaxies.  SN1998bw occurred in a spiral arm of the face-on SB galaxy
ESO~184--G82 \citep{Hol}.  The redshift of ESO~184--G82 is 0.0085
\citep{Tin}. The aims of this {\sl Letter\/} are to characterize the
region in ESO~184--G82 where SN1998bw occurred, and to measure or
constrain the brightness of SN1998bw at a late epoch more than two
years after the event.

	We use a Hubble parameter of $H_0 = 100h^{-1}$ km s$^{-1}$
Mpc$^{-1}$ and assume $\Omega_m = 0.3$ and $\Omega_{\Lambda} = 0.7$
throughout this {\sl Letter}.  For this cosmology a redshift of $z =
0.0085$ corresponds to a luminosity distance of $25.65h^{-1}$ Mpc and
a scale of $122h^{-1}$ proper parsecs per arcsecond.


\section{Observations and Data Reductions\label{obs}}

       The {\sl Hubble Space Telescope\/} ({\sl HST\/}) was used to
obtain STIS/CCD images of ESO~184--G82 between 21:47 UT and 23:42 UT
on 11 June 2000 \citep{HFT2000b}.  A log of the {\sl HST\/}
observations is given in Table~\ref{obslog}.  The total exposure times
were 1240 seconds in the 50CCD (clear, hereafter referred to as CL)
aperture and 1185 seconds in the F28X50LP (long pass, hereafter
referred to as LP) aperture\footnote{The 50CCD aperture has a central
wavelength of 5850 {\AA} and a width of 4410 {\AA}.  The F28X50LP
aperture has a central wavelength of 7230 {\AA} and a width of 2720
{\AA} (STIS Handbook v4.1)}.  The CCD gain was set to
$1\;\mathrm{e}^-/\mathrm{ADU}$ and the read-out noise was 4.46
ADU/pixel.  The data was processed through the standard STIS pipeline
and combined using the {\sc dither} (v1.2) software \citep{FH2000} as
implemented in IRAF\footnote{Image Reduction and Analysis Facility
(IRAF), a software system distributed by the National Optical
Observatories (NOAO).}  (v2.11.1)/STSDAS (v2.0.2).  We used
``pixfrac'' $= 0.6$ and a final output scale of $0\farcs0254$/pixel.
These observations were taken as part of the Cycle 9 program GO-8640:
Survey of the Host Galaxies of Gamma-Ray Bursts. The URL for the
survey is
{\tt http://www.ifa.au.dk/\~{}hst/grb\_hosts/index.html}
\citep{HFT2000a}.



\section{Results\label{SECTION:results}}

\begin{figure}
\plotone{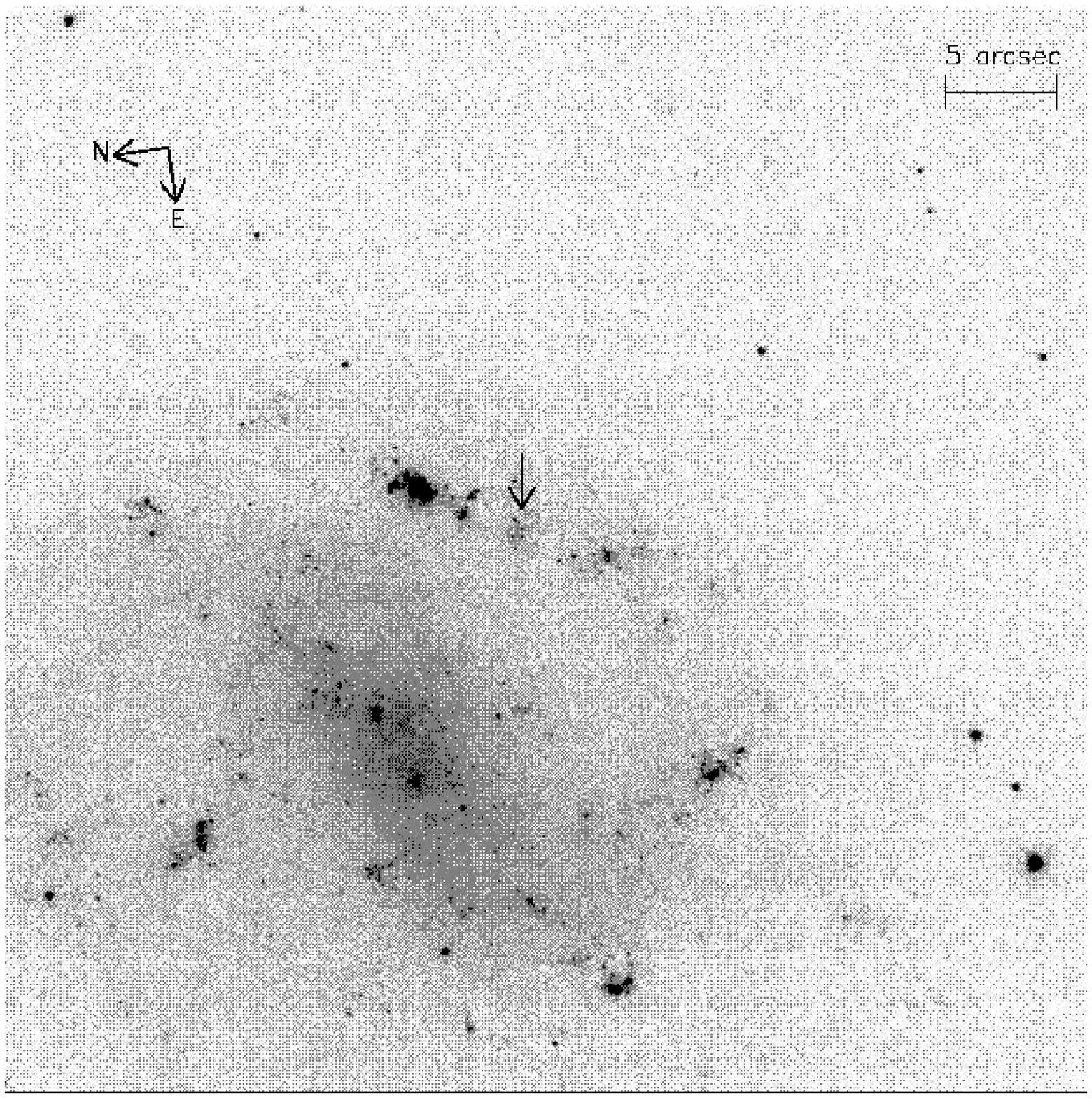}
\caption[]{The drizzled CL image. The total size of the image is
$49\arcsec \times 49\arcsec$.  The orientation of the field is shown
in the upper left corner of the image. The position angle is
$188\fdg25$ East of North. The arrow marks the position of the SN
(\S~\ref{astrometry}).\label{STISall}}
\end{figure}

	The drizzled CL image is shown as Fig.~\ref{STISall}.
ESO~184--G82 is a barred spiral galaxy of Hubble type SBc
\citep{Rau}. The magnitude of the galaxy is $B = 15.19 \pm 0.05$, $R
= 14.28
\pm 0.05$ \citep{Lau}, which corresponds to $L_B = 0.02 L^*_B$
using $M^*_B = -21.2 + 5\log_{10}(h)$ \citep{Mari}.

	The galaxy is clearly in a stage of active star formation.
Fig.~\ref{STISall} shows that the optical appearance of the galaxy is
dominated by a large number of high surface brightness star forming
regions especially in the southern spiral arm where the SN
occurred. When comparing the CL and LP images a large number of red
giants can be seen within or near the star forming regions (two
examples are seen in Fig.~\ref{STISzoom}).

	As also noted by \citet{Hol} ESO~184--G82 is member of a group
of galaxies consisting of the three galaxies ESO~184--G80, ESO~184--G81
and ESO~184--G82.  Furthermore, there is an previously unclassified Sb 
galaxy $1.\arcmin1$ south of ESO~184--G82 and a fainter bulge-dominated 
galaxy $1.\arcmin7$ south of ESO~184--G82. Hence, the star formation activity
could be enhanced by interaction (see \S~\ref{discuss}).


\subsection{Astrometry\label{astrometry}}


	The most precise determination of the absolute position of
SN1998bw on the sky is based on the radio observations made with the
Australian Telescope Compact Array April 28 1998 \citep{Wie}.  This
position is RA $=$ 19:35:03.31, Dec.\ $=$ $-$52:50:44.7 (J2000.0) with
an error of $0\farcs1$.  However, SN1998bw was very faint at the time
of the STIS observations, and there are no other sources in the STIS
Field of View (FOV) for which positions have been determined with
equally high precision. In order to determine relative astrometry
between SN1998bw and other objects in the STIS FOV we therefore
retrieved images obtained with the ESO-VLT Antu telescope with the
FORS1 instrument from the ESO archive. These images were taken at an
epoch (1999 April 18) when the SN was still clearly visible. The
journal of archive data is given in Table~\ref{archivelog}
\citep{Leib,Sol}.

	The precise location of the SN in the STIS CL image was
derived from the ground based VLT images in the following way.  The
STIS images revealed several bright blue objects within $0\farcs5$ of
the position of the SN\@.  These objects were covered under the
point-spread function (PSF) in the VLT images, which had a seeing of
about $0\farcs9$.  This affected the centroid determination of the SN
somewhat, giving rise to a colour dependent position error of $\approx
1.5$ drizzled STIS pixels ($\approx 0\farcs038$).  By smoothing,
scaling, and transforming the STIS CL and LP images, the contribution
from the host galaxy, including the clusters in the proximity of the
SN, could be subtracted completely from the VLT $V$ and $I$
images. Astrometry, based on these two subtracted images, gave
accurate and internally consistent positions, based on six and eleven
reference stars. The fits are six-parameter affine fits, with weights
calculated from a noise model. The errors are the standard deviations
of the residuals in the fit, properly normalized by weights and
degrees of freedom. The final error of the weighted position was
estimated to be about $0.7$ drizzled pixels in the CL image, or about
$0\farcs018$. In the drizzled LP the error is slightly smaller.

	In order to check this position {\sc daophot~II}
\citep{Stet,Stet2} with extensions ({\sc daomatch} and {\sc
daomaster}) was applied to the drizzled STIS LP image and the ground
based VLT $R$ image, respectively. Subsequently the two auxiliary
programs {\sc daomatch} and {\sc daomaster} were used to derive a
four-parameter transformation (offset, scale and orientation) from the
positions of five isolated stars judged to be free of superposed
nebulosities.  The transformed position of the SN is within $0.5$
pixels of the value derived from our more formal astrometric
procedure.

\subsection{The Environment of the SN\label{SNenv}}

\begin{figure}
\plotone{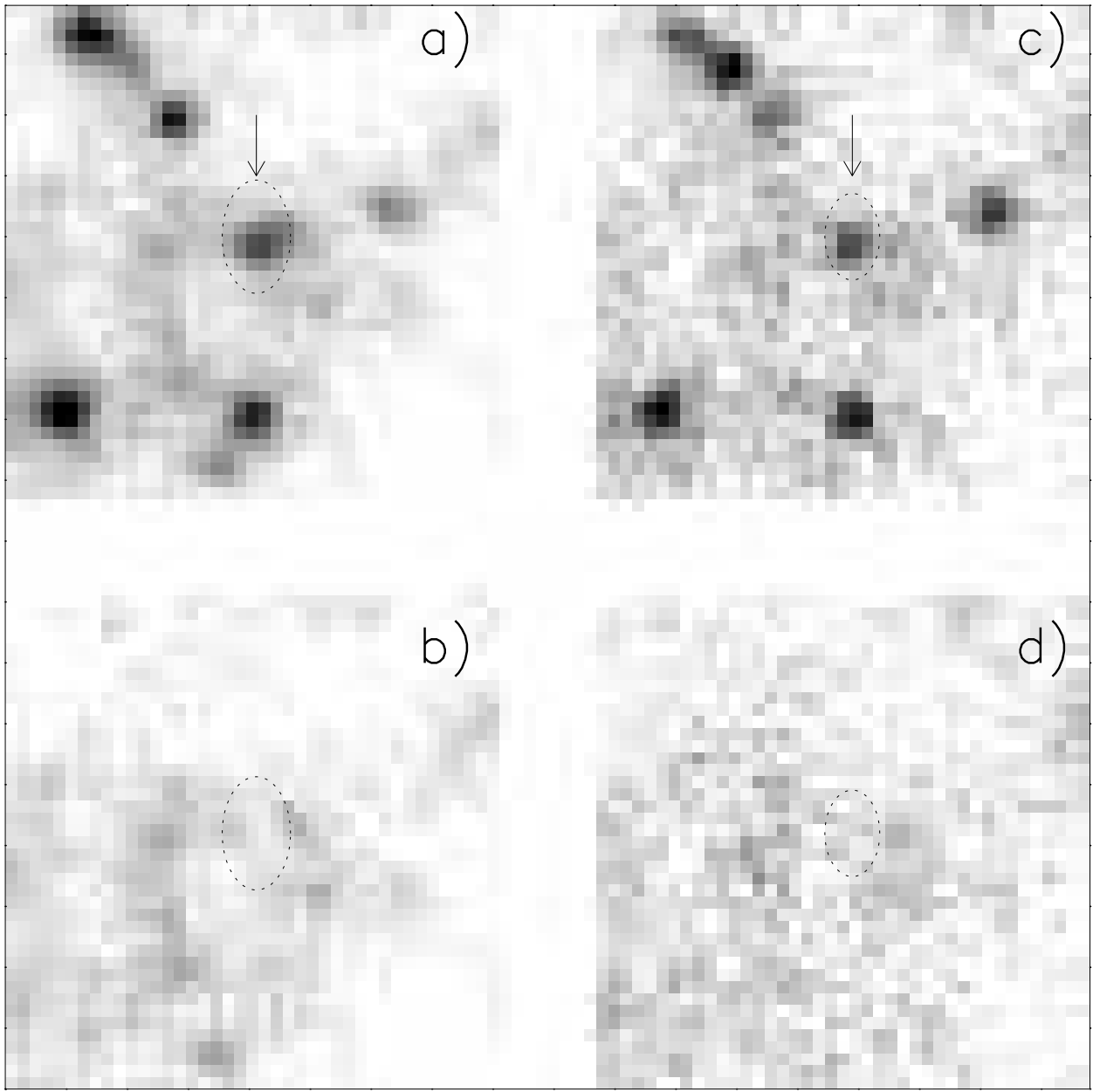}
\caption[]{The orientation of each panel is the same as if
Fig.~\ref{STISall}.  {\it a)\/} A region of $40 \times 40$ drizzled
pixels ($1\farcs0 \times 1\farcs0$) centered on the position of the SN
from the CL image. The orientation of the field is the same as in
Fig.~\ref{STISall}. The 5$\sigma$ error-ellipse for the astrometry is
shown as a dotted ellipse. The region marked by an arrow is consistent
with the position of the SN to within the astrometric error. Also seen
in the field are the six objects (s1--s6) consistent with being point
sources (see Table~\ref{tabphot}). Three of these (s2, s4, and s6)
have colours consistent with being red giant stars.  The other three
point sources have blue colours that are consistent with massive
main-sequence stars.  The is also an arc-like structure extending from
the SN towards s6.  {\it b)\/} The same region as in {\it a)\/} after
subtraction of the seven PSFs as described in
Sect.~\ref{photpsf}. {\it c) and d)\/} Same as {\it a)\/} and {\it
b)}, but for the LP image.\label{STISzoom}}
\end{figure}

	As also noted by \citet{Leib} and \citet{Sol} the SN is
superposed on an \ion{H}{2} region. The diameter of this region is
approximately between $2\arcsec$ and $3\arcsec$, corresponding to a
physical diameter of between $250h^{-1}$ and $350h^{-1}$ pc at $z =
0.0085$.  Fig.~\ref{STISzoom}a shows a region with a size of $1\farcs0
\times 1\farcs0$ centered on the position of the SN as determined in
Sect.~\ref{astrometry}.  At the exact position of the SN is the object
marked by an arrow in Fig.~\ref{STISzoom}a. The object is also seen in
the LP image shown in Fig.~\ref{STISzoom}c. In the following we will
refer to this object as the SN\@. Also seen are six objects (in the
following refered to as s1--s6) consistent with being point
sources. In the CL image there appears to be an arc-like structure
extending from the SN towards s6.

\subsection{Photometry and PSF-Subtraction}
\label{photpsf}

   We estimated the total AB magnitude of the SN, the other point
sources (s1--s6) visible in Fig.~\ref{STISzoom} and the arc-like
structure, on both the CL and LP images in the following way. For the
photometry of the point sources we use {\sc daophot~II}.  We
determined the PSF in the same way as described in \S~4 of
\citet{HH1999}.  We used {\sc allstar} \citep{SH1988} with a fitting
radius of $0\farcs071$ and a sky annulus of $0\farcs076$--$0\farcs178$
to fit a PSF to the object and determined aperture corrections by
measuring the magnitudes of several stars in apertures with radii of
$1\farcs108$ (CL) and $0\farcs963$ (LP). Tables 14.3 and 14.5 of the
STIS Instrument Handbook show that these radii correspond to $\approx
100$\% of the encircled energy for a point source.  We used the STIS
zero points given in \citep{GAR} to convert the count rates to CL and
LP magnitudes.  We transformed the CL and LP magnitudes to Johnson $V$
and Kron--Cousins $I$ magnitudes using the calibrations of
\citep{RMG2000}.  They used 33 stars with $-0.8 < V\!-\!I < 2.0$ and
found that the RMS residuals of their calibrations were $\sigma_V =
0.24$ and $\sigma_I = 0.17$.


	As seen in Fig.~\ref{STISzoom}b and Fig.~\ref{STISzoom}d all
seven objects, including the probably SN remnant in the error ellipse,
are well subtracted by the PSF\@. The results of our PSF-photometry
are given in Table~\ref{tabphot}.  The uncertainties in the $V$ and
$I$ magnitudes are dominated by the uncertainties in the
transformations from CL and LP.  The $\Delta X$ and $\Delta Y$ values
are the coordinates on the drizzled STIS CL image of each object
relative to the SN remnant. The colours of s2 and s4 are consistent
with red giant stars whereas the other four point sources have blue
colours consistent with massive main-sequence stars.

	We measured the surface brightness of the arc-like structure
in the following way.  We subtracted the point sources in
Table~\ref{tabphot} and smoothed the resulting images with a Gaussian
with a full-width at half-maximum of three pixels.  We then fit-by-eye
a line through the arc and measured the mean countrate along this
line.  The sky was estimated from empty regions of the CL and LP
images.  The results of this procedure are $\mu_V = 21.9 \pm 0.3$ and
$\mu_I = 20.3 \pm 0.3$, yielding $V\!-\!I = 1.6 \pm 0.4$.  This
structure might be the source of the H$\alpha$ emission seen in late
spectra of SN1998bw \citep{Sol}



\section{Discussion\label{discuss}}
 
\subsection{The Afterglow of GRB980425/SN1998bw}

	GRB~980425/SN1998bw occurred in the southern spiral arm of the
sub-luminous ($L_B = 0.02 L^*_B$), barred SBc galaxy
ESO~184--G82. This galaxy is in a stage of strong star formation. In
this way ESO~184--G82 resembles the LMC, which is also sub-luminous,
barred and dominated by regions with strong star formation. We note
that ESO~184--G82 is $\approx 0.3 h^{-2}$ as luminous as the LMC\@.
For $H_0 = 50$--$80$ km s$^{-1}$ Mpc$^{-1}$
\citep{dBA2000} ESO~184--G82 has a luminosity of $\approx 0.5$--$1.2
L_\mathrm{LMC}$. 

	There are two possible explanations for why ESO~184--G82 is
undergoing strong star formation. These explanations may be related.
Firstly, ESO~184--G82 has a bar.  Whereas the presence of a bar in
spiral galaxies does not, in general, correlate with the global star
formation rate \citep{K1998} the bar strength has been found to
correlate strongly with the global star formation rate of the galaxy
\citep{A1999}.  Secondly, although we only have a measured redshift
for ESO~184--G82 \citep{Tin}, the galaxy is probably member of a group
of at least five galaxies simply based on the low likelihood of a
chance projection. The nearest neighbour is an Sb galaxy at a
projected separation of only $8.3h^{-1}$ kpc. Moreover, ESO~184--G82
shows indications of being morphologically disturbed. Interaction
induced star formation is therefore a likely explanation.

  The SN occured in a star-forming region with several bright, young 
stars within a projected distance of $\sim$100 pc. In this respect,
it is similar to most core-collapse SNe that typically are
located within young stellar associations in spiral arms \citep{Barth,
VanDyk}.

	The magnitude of the emission at the position of the SN 778
days after the discovery of the GRB is $\gtrsim 1$ magnitude brighter
than one would expect from a simple radioactive decay model
\citep{Sol}. The emission in both the CL and LP image is well-fitted
by a PSF at a position that is consistent with the precise position of
the SN determined in \S~\ref{astrometry}.  The residuals remaining
after subtracting the scaled PSF are consistent with noise.  This
suggests that the majority of the observed emission could in fact be
from the SN\@.  We tested our ability to resolve a star cluster in
ESO~184--G82 by adding artificial clusters with Michie--King
\citep{M1963,K1966} profiles and the same flux as the SN remnant to
the CL image as described in \citep{HCH1999}.  We found that star
clusters with core radii of $r_c \gtrsim 0\farcs013$ ($= 1.6 h^{-1}$
pc) appeared as extended objects and are not well fit by the PSF\@.
This suggests that, if the SN did occur in a star cluster, the
underlying star cluster is either physically very small and thus can
not be distinguished from a point source, or that the underlying star
cluster is very faint.  If the emission is really dominated by the
fading SN, as can only be confirmed by follow-up observations with the
{\sl HST}, this would imply a significant flattening of the light
curve, possibly explained by interaction with circumstellar material
or accretion on a black hole \citep{Leib,PCR2000,Sol}, or the 
effect of the positron decay \citep{NMN2000}.

\subsection{Implications for GRB-Research}

	These data provide a very solid confirmation of the hypothesis
that at least some GRBs are strongly related to star formation
\citep{P1998}.  However, some GRBs, such as
GRB~970508, show evidence that there was no SN like SN1998bw
associated with them.  Therefore, either GRBs are associated with
non-standard-candle SNe, or only some GRBs are associated with SNe and
the others have a different progenitor.  \citet{GLM1999} have used
Bayesian methods to estimate that no more than $\approx 5$\% of the
GRBs that have been detected by BATSE were produced by known SNe
Ib--Ic.  However, \citep{GAL} conservatively estimate the probability
that the coincidence between GRB~980425 and SN1998bw is chance to be
$\approx 10^{-4}$.  This suggests that GRB980425/SN1998bw was a member
of an unusual class of GRBs/SNe.

        GRB~980425 had a isotropic-equivalent energy of $10^{48}$ erg
\citep{PAA2000}, making it $\approx 10^4$ less energetic than the
other GRBs for which distances (and thus isotropic-equivalent
energies) have been determined.  If GRB~980425 had occurred at $z = 1$
(the median redshift of the GRBs with measured redshifts) it would not
have been detected.  As of 30 June 2000 redshifts and total isotropic
energies have been determined for only ten GRBs, with nine having
energies between $2 \times 10^{51}$ erg and $3 \times 10^{54}$ erg,
and one having an energy of $10^{48}$ erg.  If we assume that this
ratio holds for the entire BATSE 4B Catalogue \citep{PMP1999} then the
observed number of high-energy ($E \ge 10^{51}$ erg) GRBs with known
redshifts (nine) implies that a high-energy GRB will occur at $z
\lesssim 0.01$ only once every $\approx 20\,000$ years.  Therefore, it
is not surprising that we have not observed any high-energy GRBs with
$z \lesssim 0.01$.  GRBs with isotropic-equivalent energies of
$\approx 10^{48}$ may be common at $z \gtrsim 1$, but are undetectable
with current instruments.


\acknowledgments

	We thank J. Sollerman and our anonymous referee for several 
useful comments that improved our manuscript on several important 
points. We also thank and P. Rautiainen for determining
the Hubble type of ESO~184--G82. This work was supported by
the Danish Natural Science Research Council (SNF).



\newpage

\newpage

\begin{deluxetable}{cccc}
\tablewidth{0pt}
\tablecaption{The journal of {\sl HST\/}/STIS
observations\label{obslog}}
\tablehead{%
	\colhead{Field} &
	\colhead{Date (UT)} &
	\colhead{Aperture} &
	\colhead{Exposure Time} \\
	\colhead{} &
	\colhead{} &
	\colhead{} &
	\colhead{(s)}}
\startdata
GRB 980425 & 2000 Jun 11 & 50CCD (CL)    & 1 $\times$ 60 \\
           &             &               & 4 $\times$ 295 \\
GRB 980425 & 2000 Jun 11 & F28X50LP (LP) & 3 $\times$ 296 \\
           &             &               & 1 $\times$ 297 \\
\enddata
\end{deluxetable} 

\begin{deluxetable}{ccccc}
\tablewidth{0pt}
\tablecaption{Log of FORS1 observations from the
VLT-archive\label{archivelog}}
\tablehead{%
	\colhead{Field} &
	\colhead{Date (UT)} &
	\colhead{Filter} &
	\colhead{Seeing} &
	\colhead{Exposure Time} \\
	\colhead{} &
	\colhead{} &
	\colhead{} &
	\colhead{(arcsec)} &
	\colhead{(s)}}
\startdata
GRB 980425   & 1999 Apr 18  & $V$  & 0.93 & 30  \\
GRB 980425   & 1999 Apr 18  & $R$  & 0.89 & 30  \\
GRB 980425   & 1999 Apr 18  & $I$  & 0.89 & 45  \\
\enddata
\end{deluxetable}

\begin{deluxetable}{ccccccc}
\tablewidth{0pt}
\tablecaption{Results of PSF-photometry of the objects seen in
Fig.~\ref{STISzoom}\label{tabphot}}
\tablehead{%
	\colhead{Object} &
	\colhead{$\Delta X$} &
	\colhead{$\Delta Y$} &
	\colhead{CL$_{AB}$} &
	\colhead{LP$_{AB}$} &
	\colhead{$V$} &
	\colhead{$I$}}
\startdata
SN  &  $0.0$ &   $0.0$ & 25.42$\pm$0.06 & 25.51$\pm$0.08 & 25.41$\pm$0.25 & 24.39$\pm$0.19 \\
s1 & $-14.6$ &  $17.5$ & 25.02$\pm$0.04 & 25.56$\pm$0.08 & 24.89$\pm$0.24 & 24.67$\pm$0.19 \\
s2 & $-11.2$ &  $15.4$ & 25.88$\pm$0.07 & 25.08$\pm$0.04 & 26.10$\pm$0.25 & 23.50$\pm$0.17 \\
s3 &  $-7.4$ &  $10.4$ & 25.31$\pm$0.04 & 25.70$\pm$0.10 & 25.22$\pm$0.24 & 24.73$\pm$0.20 \\
s4 &  $10.6$ &   $3.0$ & 25.90$\pm$0.06 & 25.28$\pm$0.06 & 26.07$\pm$0.25 & 23.79$\pm$0.18 \\
s5 & $-16.5$ & $-13.8$ & 24.93$\pm$0.03 & 25.06$\pm$0.05 & 24.91$\pm$0.24 & 23.96$\pm$0.18 \\
s6 &  $-0.9$ & $-14.4$ & 25.27$\pm$0.04 & 25.10$\pm$0.05 & 25.32$\pm$0.24 & 23.84$\pm$0.18 \\
\enddata
\end{deluxetable}

\end{document}